\documentclass{article}
\usepackage{spconf,amsmath,graphicx,xcolor}

\usepackage{enumitem}
\setlist{nosep, leftmargin=14pt}

\usepackage{mwe} % to get dummy images

\usepackage{xcolor}
\definecolor{paradisepink}{RGB}{233, 67, 94}

\newcommand{\RomanNumeralCaps}[1]
    {\MakeUppercase{\romannumeral #1}}

\title{A Hierarchical Transformer Encoder to Improve Entire Neoplasm Segmentation on Whole Slide Images of Hepatocellular Carcinoma}
\name{
\begin{tabular}{@{}c@{}}
Zhuxian Guo$^{\star \dagger}$ \qquad Qitong Wang$^{\star}$ \qquad Henning Müller$^{\dagger}$ \\
Themis Palpanas$^{\star \ddagger}$ \qquad Nicolas Loménie$^{\star}$ \qquad Camille Kurtz$^{\star}$
\end{tabular}}

\address{$^{\star}$ Laboratory of Informatics Paris Descartes (LIPADE), Université Paris Cité, Paris, France \\
         $^{\dagger}$ University of Applied Sciences of Western Switzerland (HES-SO Valais), Sierre, Switzerland\\
         $^{\ddagger}$ French University Institute (IUF), Paris, France}
\begin{document}
%\ninept
%
\maketitle
\begin{abstract}
In digital histopathology, entire neoplasm segmentation on Whole Slide Image (WSI) of Hepatocellular Carcinoma (HCC) plays an important role, especially as a preprocessing filter to automatically exclude healthy tissue, in histological molecular correlations mining and other downstream histopathological tasks. The segmentation task remains challenging due to HCC's inherent high-heterogeneity and the lack of dependency learning in large field of view. In this article, we propose a novel deep learning architecture with a hierarchical Transformer encoder, HiTrans, to learn the global dependencies within expanded 4096$\times$4096 WSI patches. HiTrans is designed to encode and decode the patches with larger reception fields and the learned global dependencies, compared to the state-of-the-art Fully Convolutional Neural networks (FCNN). Empirical evaluations verified that HiTrans leads to better segmentation performance by taking into account regional and global dependency information.
\end{abstract}
\section{Introduction}
\label{sec:intro}
Hepatocellular Carcinoma (HCC) is a primary tumor of the liver and is now the fifth most common cancer worldwide \cite{ferlay_cancer_2015}. HCC is a highly heterogeneous molecularly and histologically as a cancer. A series of ongoing studies have shown that the HCC phenotype appears to be closely related to particular gene mutations \cite{calderaro_molecular_2019}. 

Clustering the WSI representations of certain phenotypes to particular gene mutations by mining the relationships of the representations and their corresponding transcriptomic data is clinically meaningful. Imaging-based multi-omics can help physicians understand the morphology and micro-environmental cell population changes related to certain mutations.
Multiple Instance Learning (MIL) such as the framework Clustering-constrained Attention Multiple Instance Learning (CLAM) \cite{lu_data-efficient_2021} has been applied in \cite{zeng_artificial_2022} to predict the activation of 6 common HCC immune gene signatures within a roughly selected neoplasm area. 
A robust automatic neoplasm segmentation can then act as a preprocessing filter, not only to produce the annotation in lieu of pathologists but also to exclude the potential annotation bias to specific highly predictive regions when using the annotations provided by experienced pathologists.

Automatic neoplasm segmentation remains an important challenge today mainly due to: 
(1) the inherently high tissue heterogeneity, and (2) the lack of consideration of effectively aggregated large-field relational features. 
For instance, HCC is a highly heterogeneous cancer and it has distinct morphological phenotypes. The morphological appearance of different phenotypes is very different across cases, which makes the tumor segmentation model hard to generalize.

The \textit{de facto} WSI segmentation models are patch-based convolutional neural networks (CNN) like \cite{DBLP:journals/eswa/TorresMGG20,roy_convolutional_2021}, and the patch size usually ranges from 128$\times$128 to 512$\times$512 due to GPU memory limitations. Such a patch size is very small compared to the gigapixel size of the WSI, which limits the receptive field of the model and leads to a limited ability to capture large-scale aberrant tissue structures in HCC such as macro-trabeculae\footnote{Neoplastic cells of macrotrabecular-massive HCC are arranged in thick trabeculae surrounded by vascular spaces.}, pseudoglandular and necrotic foci architectural patterns, etc.

Previous work~\cite{DBLP:journals/mia/SchmitzMNKSWR21} has studied how the proposed multi-scale CNN models take into account the histological features at different scales, ranging from nuclear aberrations through cellular structures to the global tissue architecture, by using patches at different spatial scales as input. 
For the multi-scale CNN model, larger-scale patches, which are centered aligned to the smallest scale patch at full resolution (detail patch), are downsampled to similar size to the detail patch in order to fit the segmentation framework.
A general image down-sampler is usually applied while the WSI complexity is quite different from other types of images. 
In this context, we aim to develop an entire HCC neoplasm segmentation framework by using state-of-the-art (SOTA) approaches to mimic the WSI exploration by pathologists in a hierarchical fashion and thus to serve for downstream tasks in digital histopathology.

Transformers \cite{DBLP:conf/nips/VaswaniSPUJGKP17} rely on global self-attention mechanisms and have achieved excellent performance in many tasks with global dependency requirements such as sequence modeling and language modeling. 
They were modified for computer vision tasks, called Vision Transformers (ViT) to serve as an alternative to CNNs in feature extraction \cite{DBLP:conf/iclr/DosovitskiyB0WZ21}. 
In digital pathology, as a follow up of CLAM, ViT was shown to be stronger in feature aggregation than simpler attention-weighted average mechanisms. They can also be stacked as a hierarchical architecture to effectively aggregate the WSI features to a slide-level representation \cite{DBLP:conf/cvpr/ChenC0CTKM22}. 

Inspired by the success of the application of ViT on representation learning of gigapixel images, we propose in this article a framework, called HiTrans, with hierarchical-based Transformer encoder to enlarge the field of view and to enhance the entire HCC neoplasm segmentation.
Such a contribution allows to dramatically increase the segmentation field of view to 4096$\times$4096. 
The WSI patches are encoded with larger fields of view compared to conventional Fully Convolutional Neural networks (FCNN), and are decoded by taking into account regional and global dependency information. 
The experimental results with a large real dataset demonstrate that the proposed HiTrans framework can lead to better entire HCC neoplasm segmentation, quantitatively and qualitatively.

The dataset used in this study is introduced in Sec.~\ref{sec:material}. 
Sec.~\ref{sec:method} presents the data preprocessing pipeline, the baseline architecture, and the proposed network training protocol. 
Experimental results are provided in Sec.~\ref{sec:results}.
\section{Dataset}
% Material}
\label{sec:material}
The PAIP liver cancer segmentation challenge was held in 2019 (PAIP 2019) \cite{DBLP:journals/mia/KimJLPMHPLKHJLR21} as part of the MICCAI 2019 Grand Challenge for Pathology. The PAIP 2019 training cohort consists of 50 anonymized WSIs at the 20$\times$ magnification in ScanScop Virtual Slide (SVS) format. Each WSI was selected from the HCC resection slides from one patient, which means the 50 WSIs in the training cohort belong to 50 different individuals. 
The Edmonson-Steiner tumor grade distribution is 7, 23, 20 for Grade I, II, III, respectively.
The slides were all stained with conventional hematoxylin and eosin (H\&E) staining and were digitized with an Aperio AT2 whole-slide scanner. 
The WSI size ranges from 35855$\times$39407 to 64768$\times$47009. 
The training cohort WSIs come with two-layers of annotation for whole tumor areas and viable tumor areas. 
Only the first annotation layer (i.e., the whole tumor area) was used in this study. The whole tumor area means that the entire neoplasm that can be observed on the WSI, including all dispersed viable tumor cell nests, tumor necrosis and tumor capsules.
\section{Methods}
\label{sec:method}
The proposed HiTrans framework (Fig.~\ref{fig:res}) takes 4096$\times$4096 WSI patches as input. 
A hierarchical Transformer encoder add-on module is added between a ResNet \cite{DBLP:conf/cvpr/HeZRS16} encoder backbone and a modified U-Net decoder to learn the global dependencies (red dashed box). Sec.~\ref{ssec:prepro} introduces the data preprocessing pipeline. The proposed architecture details are illustrated in Sec.~\ref{ssec:archi} and the training protocol is described in Sec.~\ref{ssec:training}.
\subsection{Data preprocessing}
\label{ssec:prepro}
Since the WSI tissue mask is not provided, we followed a conventional pipeline to patchify WSIs to create the pairs of high tissue percentage 4096$\times$4096 patches and their corresponding neoplasm masks.
The 50 WSIs were split into 30, 10, and 10 for training, validation, and test, respectively.
\subsection{Proposed architecture}
\label{ssec:archi}
\begin{figure*}[htb!]
\begin{minipage}[b]{1.0\linewidth}
  \centering
  \centerline{\includegraphics[width=13cm]{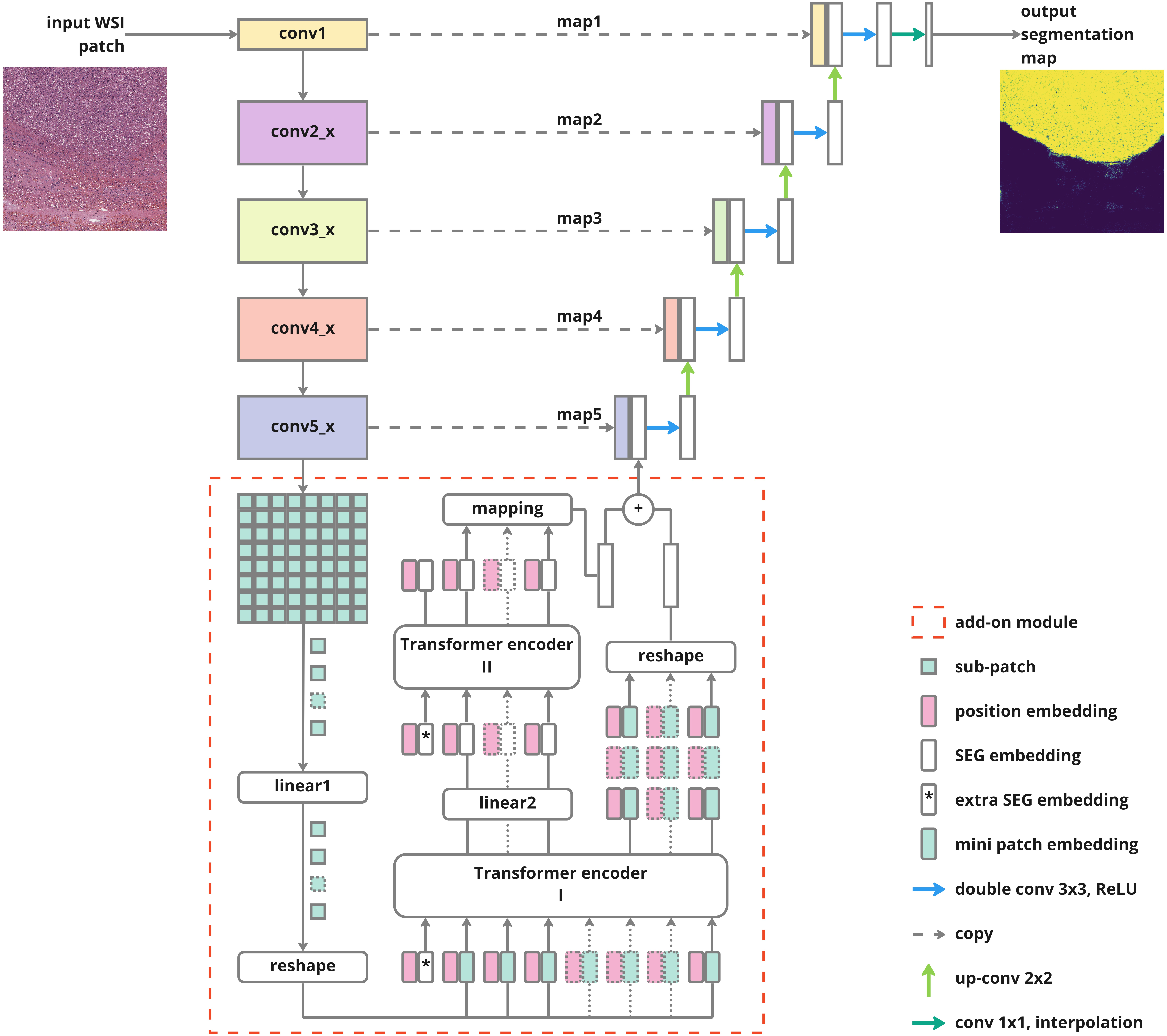}}
%  \vspace{2.0cm}
%  \centerline{ssss}\medskip
\end{minipage}
\caption{Proposed hierarchy-based Transformer encoder architecture (HiTrans) for entire HCC neoplasm segmentation.}
\label{fig:res}
\end{figure*}

A hierarchical add-on Transformer encoder module that contains two Transformers is added between the CNN feature extractor and the decoder to learn subtle global dependencies, as shown in Fig.~\ref{fig:res} the red dashed box.

\noindent {\bf [CNN feature extractor]}
The intermediate layers of a pretrained 18-layer ResNet were used as an encoder for feature extraction. The aforementioned ResNet was pretrained on 57 histopathology datasets in a self-supervised learning fashion \cite{CIGA2022100198} following the SimCLR \cite{DBLP:conf/icml/ChenK0H20} contrastive learning setting. The adaptive average pooling layer and the dense layer at the end of the ResNet were removed to make it as a 2D feature map extractor. 
Five feature maps are generated (Fig.~\ref{fig:res}, map1 to map5).

% \subsubsection{
\noindent {\bf [Transformer encoder \RomanNumeralCaps{1}]}
Transformer encoder \RomanNumeralCaps{1} is a 12-layer standard Transformer encoder with six attention heads and 384-length hidden dimension.
Map5 is unfolded into 64 (8$\times$8) seamless 16$\times$16 sub-patches, and each sub-patch contains 256 mini patch embeddings. The sub-patches are linearly transformed (Fig.~\ref{fig:res}, linear1) to fit the hidden dimension of Transformer encoder \RomanNumeralCaps{1}. An extra segmentation embedding SEG is added. All output SEG embeddings of each sub-patch that contain the regional features of each sub-patch are kept to form the inputs of Transformer encoder \RomanNumeralCaps{2}. All output mini patch embeddings are reshaped to a 128$\times$128 feature map to act as a skip connection to provide finer regional features for decoding.

% \subsubsection{
\noindent {\bf [Transformer encoder \RomanNumeralCaps{2}]}
Transformer encoder \RomanNumeralCaps{2} has 12 layers with three attention heads, and the hidden dimension is 192. It takes the output SEG embeddings from Transformer encoder \RomanNumeralCaps{1} to learn the global dependencies among the sub-patch features, within a 4096$\times$4096 patch. The output embeddings of the sub-patches are reshaped to a 8$\times$8 feature map. 
Thanks to the self attention mechanism, each element in the feature map contains its global dependency information. 
Each element is then mapped and added up to its spatial corresponding elements on the 128$\times$128 feature map from Transformer encoder \RomanNumeralCaps{1}. We added up the feature maps from the two Transformer encoders instead of doing concatenation in order to alleviate the block-biased prediction.

% \subsubsection{
\noindent {\bf [Global dependency learning]}
The feature maps from the two Transformer encoders are fused to merge regional and global features. The new feature map is then concatenated with map5 to maintain localization accuracy. Performing a global dependency learning through this hierarchical Transformer encoder architecture, the WSI patches are encoded with larger fields of view compared to CNN. The proposed architecture also provides the decoder with regional and global dependency information for a finer segmentation.

\noindent {\bf [Convolutional decoder with shortcuts]}
Transposed convolutional layers (Fig.~\ref{fig:res}, up-conv 2$\times$2) expand the feature map size. Map1 to map4 are concatenated with the expanded feature maps of the $l-1$ layers, and then pass the double convolutional layers that halve the number of channels. 
At the end, the 2048$\times$2048 feature maps will be decoded to a  4096$\times$4096 segmentation map by a $1\times1$ convolutional layer following a bilinear interpolation with corner pixels alignment.

\subsection{Network training}
\label{ssec:training}
% \subsubsection{
\noindent {\bf [General training protocol]}
The model was trained on a Nvidia A100 SXM4 80GB GPU for 100 epochs using the AdamW optimizer \cite{DBLP:journals/corr/abs-1711-05101} with a batch size of 2 and a early stopping patience of 10. 
Base learning rate was set to 5e-4, with the first 10 epochs used to warm up followed by decay using a cosine schedule to reduce the base learning rate to the minimum learning rate 1e-6. Weight decay rate was increased gradually from 1e-2 to 1e-4 following a cosine schedule.

% \subsubsection{
\noindent {\bf [Alternate training]}
Alternate training strategy was adopted to overcome the convergence difficulty in training this hierarchical semantic segmentation architecture with two stacked Transformer encoders. The ResNet feature extractor, Transformer encoder \RomanNumeralCaps{1} and \RomanNumeralCaps{2} were trained Alternately to maximize the framework ability.

\section{Results}
\label{sec:results}

\subsection{Evaluation metric and results}
The WSI segmentation for all models was performed in seamless patch-wised inference units. 
The average Jaccard index of the 10 WSIs in the test set is used as a quantitative score to evaluate the entire neoplasm segmentation performance.
Notably, the results we presented cannot be directly compared with PAIP 2019 leaderboard results, because we focused on evaluating the model performance and avoided the usage of postprecessing steps, including manually-designed WSI postprocesing strategies, ensemble learning, and overlapped inference.
Moreover, the results in this study is only trained and tested on the training cohort of PAIP 2019 and only the whole tumor area annotation is used for training.

Compared with three SOTA semantic segmentation frameworks, U-Net \cite{DBLP:conf/miccai/RonnebergerFB15}, DeepLabV3 \cite{DBLP:journals/corr/ChenPSA17}, PSPNet \cite{DBLP:conf/cvpr/ZhaoSQWJ17}, and one SOTA Transformer-based framework, SegFormer \cite{DBLP:conf/nips/XieWYAAL21}, the proposed HiTrans framework has better performance on HCC segmentation task (Tab.~\ref{table1}, Exp.~8). Among the FCNN segmentation frameworks, DeepLabV3 with a ResNet-50 backbone has the best performance (Tab.~\ref{table1}, Exp.~3) thanks to the stronger feature extractor and the Atrous Spatial Pyramid Pooling (ASPP) modules, which probe convolutional features at multiple scales while avoiding increasing network size too much. For SegFormer, the smaller variant SegFormer-B0 (Tab.~\ref{table1}, Exp.~6) has better performance than the larger variant SegFormer-B2 (Tab.~\ref{table1}, Exp.~7), possibly due to the convergence difficulty in training larger Transformer-based model. 

\begin{table}[t!]
 \begin{center}
 \begin{tabular}{ccccc}
 \hline
 Exp.~& Method & Patch size & Avg. Jaccard \\
 \hline
 1 & U-Net & 512 & 0.6659 \\
 \hline
 2 & R18\_DLv3 & 512 & 0.6272 \\
 3 & R50\_DLv3 & 512 & 0.6803 \\
 \hline
 4 & R18\_PSPNet & 512 & 0.6252 \\
 5 & R50\_PSPNet & 512 & 0.6695\\
 \hline
 6 & SegFormer-B0 & 512 & 0.6352 \\
 7 & SegFormer-B2 & 512 & 0.6027 \\
 \hline
 8 & HiTrans & 4096 & \textbf{0.7513}\\
 \hline
 \end{tabular}
 \caption{Experimental results and transversal comparison. \label{table1}}
 \end{center}
\end{table}

\begin{table}[t!]
 \begin{center}
 \begin{tabular}{ccccc}
 \hline
Exp.~& Method & Patch size & Avg. Jaccard \\
 \hline
 1 & R18\_U-Net & 512 & 0.6609 \\
 2 & R18\_U-Net & 4096 & 0.7202 \\
 3 & TR-\RomanNumeralCaps{1} & 4096 & 0.7172\\
 4 & HiTrans & 4096 & \textbf{0.7513}\\
 \hline
 \end{tabular}
 \caption{Quantitative results of the ablation study. \label{table2}}
 \end{center}
\end{table}

\begin{figure}[htb!]
\begin{minipage}[b]{.48\linewidth}
  \centering
  \centerline{\includegraphics[width=4.0cm]{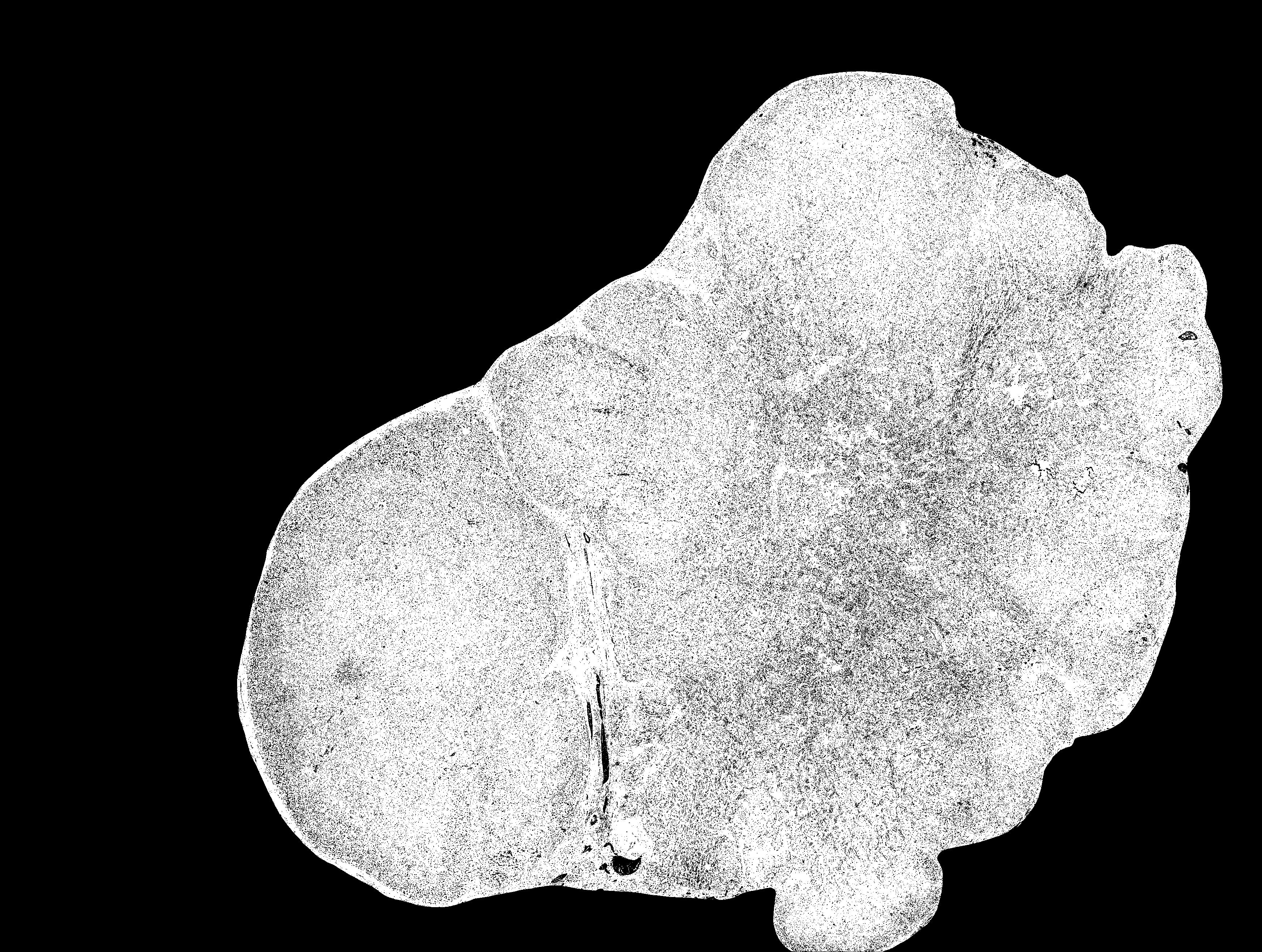}}
%  \vspace{1.5cm}
  \centerline{(a) Ground truth}\medskip
\end{minipage}
\hfill
\begin{minipage}[b]{.48\linewidth}
  \centering
  \centerline{\includegraphics[width=4.0cm]{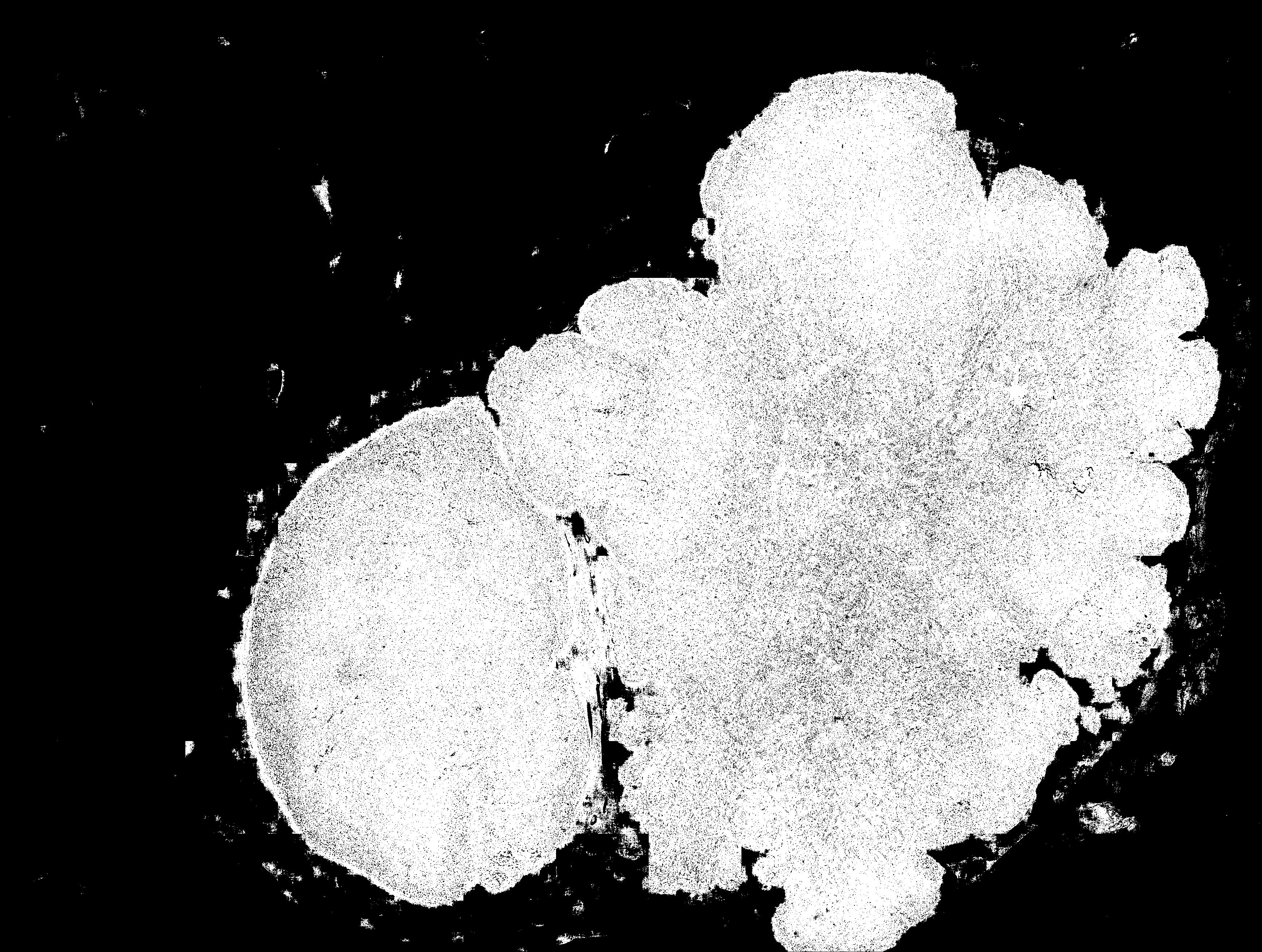}}
%  \vspace{1.5cm}
  \centerline{(b) R18\_U-Net (4096)}\medskip
\end{minipage}
\begin{minipage}[b]{.48\linewidth}
  \centering
  \centerline{\includegraphics[width=4.0cm]{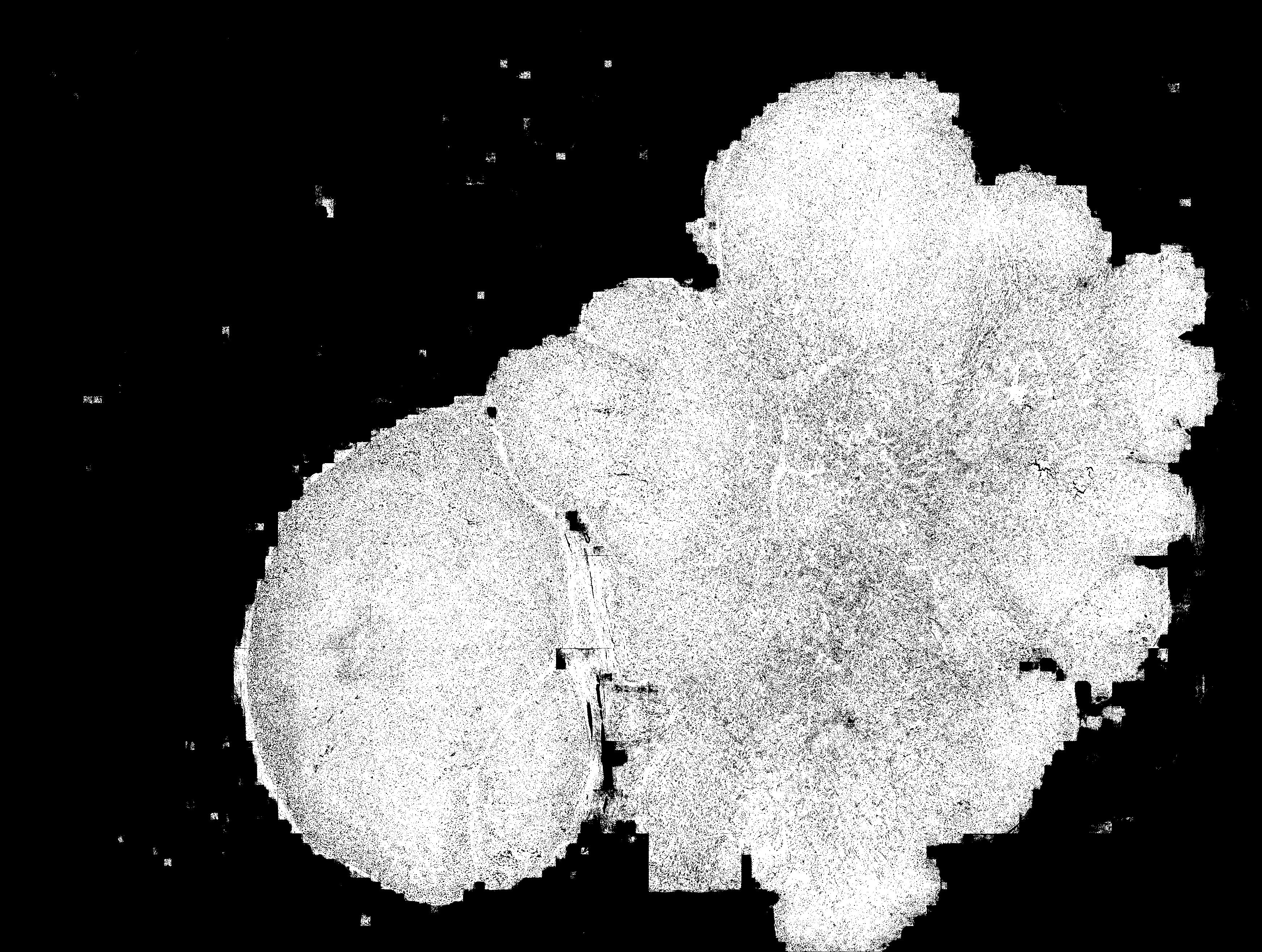}}
%  \vspace{1.5cm}
  \centerline{(c) TR-\RomanNumeralCaps{1}}\medskip
\end{minipage}
\hfill
\begin{minipage}[b]{.48\linewidth}
  \centering
  \centerline{\includegraphics[width=4.0cm]{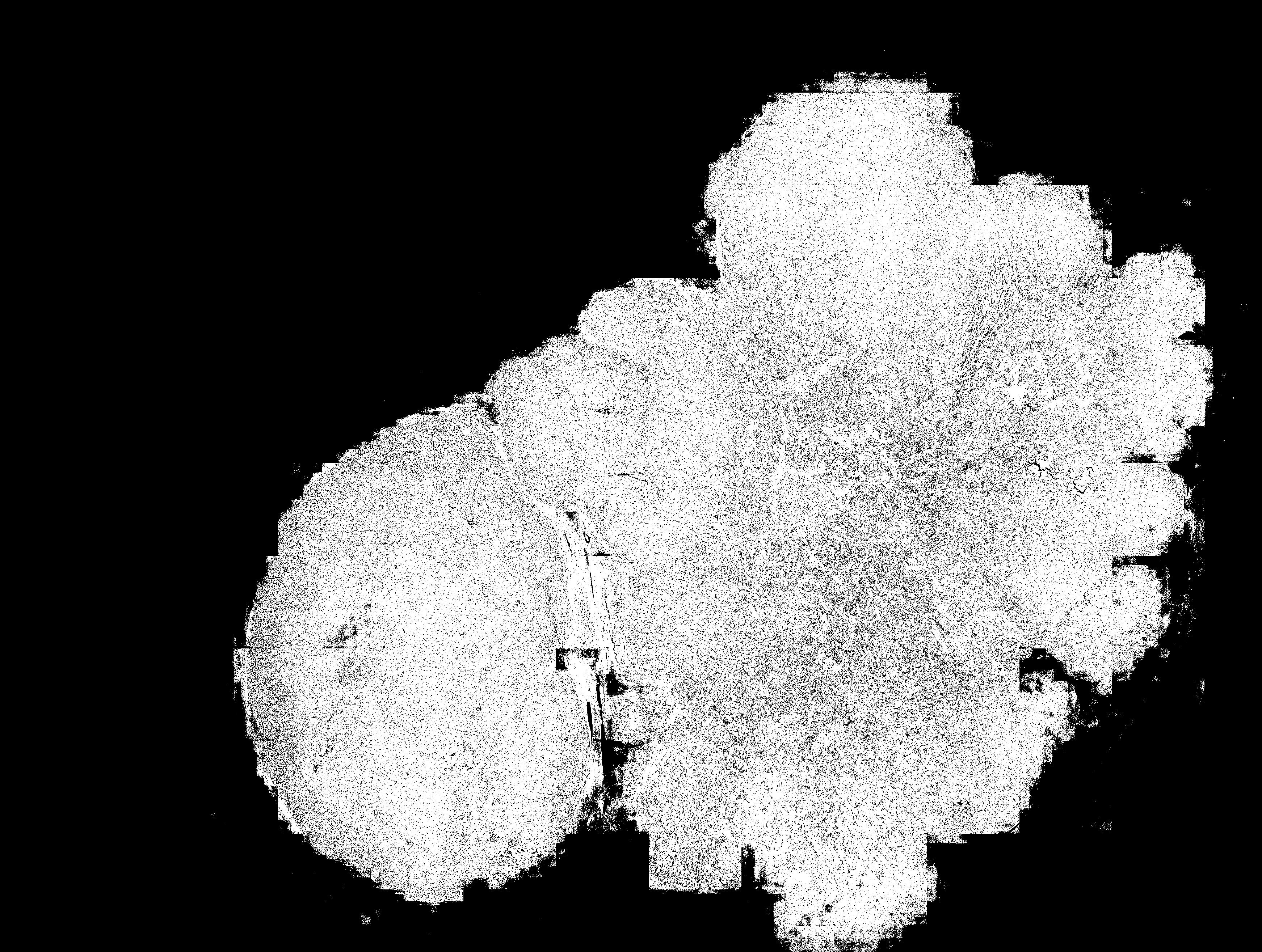}}
%  \vspace{1.5cm}
  \centerline{(d) HiTrans}\medskip
\end{minipage}
\caption{Qualitative segmentation results of the ablation study.}
\label{fig:vis}
\end{figure}

\subsection{Ablation study}
The proposed hierarchical Tansformer encoder framework can learn global dependencies and bring muti-scale cues for the decoder during segmentation inference. In the ablation study, we removed the Transformer hierarchical add-on module, in order to evaluate the above argument, namely, that knowledge of global dependencies can enhance segmentation performance. Experiments without the add-on module using 512$\times$512 (Tab.~\ref{table2}, Exp.~1) and 4096$\times$4096 (Tab.~\ref{table2}, Exp.~2) patches were conducted separately.
Besides, an experiment using an architecture without Transformer encoder \RomanNumeralCaps{2} was conducted (Tab.~\ref{table2}, Exp.~3). By comparing the experimental results, taking larger patches as input and using HiTrans to learn the global dependencies can lead to better segmentation results. Comparing with the results from add-on module dropped (Fig.~\ref{fig:vis}, b) and Transformer encoder \RomanNumeralCaps{2} dropped (Fig.~\ref{fig:vis}, c) architecture, HiTrans can further improve the precision (Fig.~\ref{fig:vis}, d) thanks to this regional and global dependency-aware architecture.

\section{Conclusions}
\label{sec:conclu}

In this article, we introduce a hierarchical Transformer-based segmentation architecture, HiTrans, for HCC entire neoplasm segmentation. HiTrans can efficiently learn the regional and global dependencies within 4096$\times$4096 WSI patches by encoding and decoding the WSI in a hierarchical fashion. The experimental results with a large real dataset demonstrate that HiTrans can lead to quantitatively and qualitatively better entire HCC neoplasm segmentation.
In our future studies, we aim at developing a robust slide-wise context aware framework by leveraging different strategies in global dependency learning like graph-based neural networks. 
We will also explore the application on other tasks.

\section{Compliance with ethical standards}
\label{sec:ethics}

This research study was conducted retrospectively using human subject data made available in open access by PAIP 2019 Challenge. Ethical approval was not required as confirmed by the license attached with the open access data.

\section{Acknowledgments}
\label{sec:acknowledgments}

This work was supported by Data Intelligence Institute of Paris (diiP), IdEx Universit\'e Paris Cit\'e (ANR-18-IDEX-0001), and Translational Research Program in Cancerology INCa-DGOS - PRTK-2020, and was performed using HPC resources from GENCI-IDRIS (2022-AD011012825R1) made by GENCI. 
Qitong Wang is funded by China Scholarship Council.


\bibliographystyle{IEEEbib}
\bibliography{strings,refs, references}

\end{document}